\documentclass[journal,twoside,web]{ieeecolor}

\usepackage{cite}
\usepackage{xcolor}
\usepackage{amsmath,amssymb,amsfonts,mathtools}
\usepackage{graphicx}
\usepackage{booktabs}
\usepackage{textcomp}
\usepackage{lcsys}
\usepackage{threeparttable}
\usepackage{url}

\usepackage{tikz}
\usetikzlibrary{arrows}

\graphicspath{{./fig/}}

\DeclareMathOperator{\atantwo}{atan2}
\DeclareMathOperator{\sgn}{sgn}
\newcommand{\Real}{\operatorname{Re}}
\newcommand{\Lie}{\mathcal{L}_{\boldsymbol{f}}}
\newcommand{\bx}{\boldsymbol{x}}
\newcommand{\bg}{\boldsymbol{g}}
\newcommand{\bff}{\boldsymbol{f}}
\newcommand{\bxi}{\boldsymbol{\xi}}
\newcommand{\bl}{\boldsymbol{\ell}}

\newtheorem{theorem}{Theorem}[section]
\newtheorem{lemma}[theorem]{Lemma}
\newtheorem{proposition}[theorem]{Proposition}

\newtheorem{remark}[theorem]{Remark}

\begin{document}

\def\BibTeX{{\rm B\kern-.05em{\sc i\kern-.025em b}\kern-.08em
    T\kern-.1667em\lower.7ex\hbox{E}\kern-.125emX}}
\markboth{\journalname, VOL. XX, NO. XX, XXXX 2017}
{Author \MakeLowercase{\textit{et al.}}: Preparation of Papers for IEEE Control Systems Letters (August 2022)}

\title{Learn Locally, Extend Globally: Koopman Eigenfunctions and the Basin of Attraction}

\author{
Marcos~Netto,~\IEEEmembership{Senior~Member,~IEEE}, Aditya~Kale,~\IEEEmembership{Student~Member,~IEEE}, Bukunmi~G.~Odunlami,~\IEEEmembership{Student~Member,~IEEE}, Andrey~Bernstein,~\IEEEmembership{Member,~IEEE}
\thanks{This work is supported in part by Eaton Research Laboratories and in part by the National Science Foundation under Grant No. 2328241. M. Netto, A. Kale, and B. G. Odunlami are with the Department of Electrical and Computer Engineering, New Jersey Institute of Technology, Newark, NJ 07102, USA. A. Bernstein is with Eaton Research Laboratories, Golden, CO 80401, USA.}
}

\maketitle
\thispagestyle{empty}

%%%%%%%%%%%%%%%%%%%%%%%%%%%%%%%%%%%%%%%%%%%%%%%%%%%%%%%%%%%%%%%%%%%%%%%%%%%%%%%%
\begin{abstract}
Koopman eigenfunctions associated with the Jacobian eigenvalues of a stable equilibrium are Lyapunov functions defined exactly on its basin of attraction. We turn this into a computational procedure for power systems. Eigenfunctions estimated by extended dynamic mode decomposition (EDMD) are written as functions of the state, made to vanish at the equilibrium by centering the lifted variable $\cos\delta$, and verified on a grid with the model: two Lyapunov candidates from the EDMD spectrum certify up to 66\% of the basin of a single-machine infinite-bus system from data in the linear region. The Koopman semigroup property then extends the EDMD-Lie eigenfunction to the whole basin, recovered at the grid resolution with its isostables and isochrons.
\end{abstract}


\section{Introduction}

\IEEEPARstart{T}{he} basin of attraction of a stable equilibrium determines which post-fault states of a power system return to synchronous operation, and direct methods for transient stability assessment rest on estimates of it \cite{Chiang1987}. The classical route is a Lyapunov function, constructed algorithmically \cite{Anghel2013} and then either enlarged by integrating a certified set backward in time \cite{Genesio1985} or checked on samples with Lipschitz margins \cite{Cavraro2026}. Koopman eigenfunctions tied to the Jacobian eigenvalues of the equilibrium \cite{Mauroy2020} exist on the whole basin, uniquely under nonresonance \cite{Kvalheim2021}. They are Lyapunov functions there \cite{Mauroy2016}, with the isostables as level sets \cite{Mauroy2013}.

% fixed points where nearby trajectories are attracted along one direction (stable manifold) and repelled along another (unstable manifold).

Extended dynamic mode decomposition (EDMD) estimates eigenfunctions from data snapshots \cite{Williams2015, Korda2018}, but the estimate is only as good as the dictionary \cite{Netto2021}, and its accuracy fades with the distance from the data. Some authors therefore build the candidate eigenfunctions from the model: Bierwart and Mauroy validate it afterward by sum-of-squares programs or an adaptive grid \cite{Bierwart2025}, Umathe and Vaidya \cite{Umathe2024} work with the eigenfunctions of the saddles, and R{\"o}mer and Breitenhuber fit a rational ansatz to the Koopman generator equation through a non-convex program \cite{Romer2024}. Others keep the eigenfunction as learned: Deka and Dimarogonas sharpen it with Laplace averages before learning a Lyapunov function from it \cite{Deka2023}, and Matavalam \emph{et al.} \cite{Matavalam2024} apply it directly for transient stability assessment.

This letter certifies the learned eigenfunction itself. The left eigenvectors of the EDMD matrix give an eigenfunction in closed form, so it can be checked against the model at every grid point with an explicit tolerance. Once certified, it extends to the whole basin of a single-machine infinite-bus system. An EDMD eigenfunction with discrete eigenvalue $\mu_k \ne 1$ vanishes at the fixed point whenever the lifted fixed point is invariant under the EDMD matrix. We show that centering $\cos\delta$ ensures that invariance for any data set and removes the spurious eigenvalue of \cite{Netto2021}. Two Lyapunov candidates from the EDMD spectrum, certified on a grid with the model, give inner basin estimates for five dictionaries. The semigroup property extends the eigenfunction from a small certified seed to the whole basin; the extension is finite exactly on the basin and yields the isostables and isochrons \cite{Mauroy2013}. The extension accepts any certified positively invariant seed, including a ball from the linearization. Seeding it with the eigenfunction learned from the analytical dictionary of \cite{Netto2021}, tailored to the model, keeps the object that serves state estimation. It also keeps the dictionary at six observables, since the seed need not be large.

\section{Preliminaries}
Consider $\dot{\bx}=\bff(\bx)$ on $\mathbb{X}\subseteq\mathbb{R}^n$ with flow $S_t$. The Koopman semigroup $\mathcal{K}_t$ acts on observables $g:\mathbb{X}\to\mathbb{C}$ by composition, $\mathcal{K}_t g = g\circ S_t$ \cite{Mauroy2020}. A Koopman eigenfunction with eigenvalue $\lambda\in\mathbb{C}$ is an observable $\phi:\mathcal{D}\to\mathbb{C}$, defined on a positively invariant open set $\mathcal{D}\subseteq\mathbb{X}$ and continuously differentiable there, that satisfies
\begin{align}
\phi(S_t\bx) &= e^{\lambda t}\phi(\bx), &\qquad\text{(semigroup form)}\label{eq.efun1} \\
\Lie\phi \coloneqq \nabla\phi\cdot\bff &= \lambda\phi, &\qquad\text{(generator form)}\label{eq.efun2}
\end{align}
for all $\bx\in\mathcal{D}$ and $t\ge 0$, where $\nabla$ denotes the vector differential operator and $\Lie$ denotes the Lie derivative, the semigroup generator. Eqs. \eqref{eq.efun1}--\eqref{eq.efun2} state the eigenfunction property in two equivalent forms, the semigroup form in \cite{Mauroy2020} and the generator form in \cite{Mauroy2016}. Let $\bx^{\star}$ be a hyperbolic stable equilibrium (no Jacobian eigenvalue on the imaginary axis) with basin $\mathcal{B}(\bx^{\star})$ and Jacobian eigenvalues $\lambda_1,\dots,\lambda_n$ with left eigenvectors $\bl_1^{\mathsf{T}},\dots,\bl_n^{\mathsf{T}}$. Under non-resonance conditions, the $\lambda_k$, $k=1,...,n$, are Koopman eigenvalues. The associated eigenfunctions $\phi_k$, called principal, are $C^1$ on $\mathcal{B}(\bx^{\star})$, with $\mathcal{D}=\mathcal{B}(\bx^{\star})$, satisfy $\nabla\phi_k(\bx^{\star})=\bl_k^{\mathsf{T}}$, and vanish only at $\bx^{\star}$ \cite{Kvalheim2021, Mauroy2016}. Let $\bx_0$ denote $\bx$ at $t=0$ and $\lambda_k=\sigma_k+j\nu_k$. Along a trajectory $\bx(t)=S_t\bx_0$, the semigroup form \eqref{eq.efun1} yields
\begin{align*}
\phi_k(\bx(t)) &= e^{\lambda_k t}\phi_k(\bx_0), \\
|\phi_k(\bx(t))| &= |e^{\lambda_k t}|\,|\phi_k(\bx_0)| = |e^{\sigma_k t}|\,|e^{j\nu_k t}|\,|\phi_k(\bx_0)| \\
&= e^{\Real(\lambda_k) t}|\phi_k(\bx_0)|
\end{align*}
since $|e^{j\nu_k t}|=1$, and differentiating in $t$,
\begin{align}
\tfrac{d}{dt}\,|\phi_k(\bx(t))| &= \Real(\lambda_k)e^{\Real(\lambda_k) t}|\phi_k(\bx_0)| \nonumber\\
&= \Real(\lambda_k)\,|\phi_k(\bx(t))| < 0.
\label{eq.decay}
\end{align}
The inequality holds away from $\bx^{\star}$ because $\Real(\lambda_k)<0$ at a hyperbolic stable equilibrium and $\phi_k$ vanishes only at $\bx^{\star}$. Hence, $|\phi_k|$ is a Lyapunov function on the basin. Products $\phi_k\phi_l$ are again eigenfunctions, with eigenvalue $\lambda_k+\lambda_l$, and $\phi_1^*$ is an eigenfunction with eigenvalue $\lambda_1^*$ because $\bff$ is real. Thus, $|\phi_1|^2=\phi_1\phi_1^{*}$ is a real eigenfunction with eigenvalue $\lambda_1+\lambda_1^*\!=\!2\Real(\lambda_1)$. The basin is the largest set on which the principal eigenfunctions are $C^1$ \cite{Mauroy2016}. Level sets of $|\phi_1|$ are the isostables and level sets of $\arg\phi_1$ are the isochrons \cite{Mauroy2013}.

\subsection{Extended dynamic mode decomposition}
Given snapshot pairs $\{\bx_{k-1},\bx_k\}$, $k=1,\dots,m$, sampled with period $\Delta t$, and a vector of $q$ observables $\bg(\bx)$, EDMD forms $\boldsymbol{G}_0\!=\![\bg(\bx_0)\cdots\bg(\bx_{m-1})]$, $\boldsymbol{G}_1\!=\![\bg(\bx_1)\cdots\bg(\bx_m)]$ and
\begin{equation}
\boldsymbol{K}=\boldsymbol{G}_1\boldsymbol{G}_0^{\dagger}\in\mathbb{R}^{q\times q},
\label{eq.edmd}
\end{equation}
with $\dagger$ the Moore--Penrose pseudoinverse \cite{Williams2015}. If $\mu_k$ and $\bxi_k^{\mathsf{T}}$ are the eigenvalues and left eigenvectors of $\boldsymbol{K}$, then $\hat\lambda_k=\ln(\mu_k)/\Delta t$ approximates a Koopman eigenvalue and
\begin{equation}
\hat\phi_k(\bx) = \bxi_k^{\mathsf{T}}\bg(\bx) = \sum_{i}\xi_{k,i}\,g_{i}(\bx)
\label{eq.efun_edmd}
\end{equation}
approximates the eigenfunction, as the projection of \eqref{eq.efun2} onto the span of the observables dictionary \cite{Korda2018}.

\subsection{Test system and its basin}
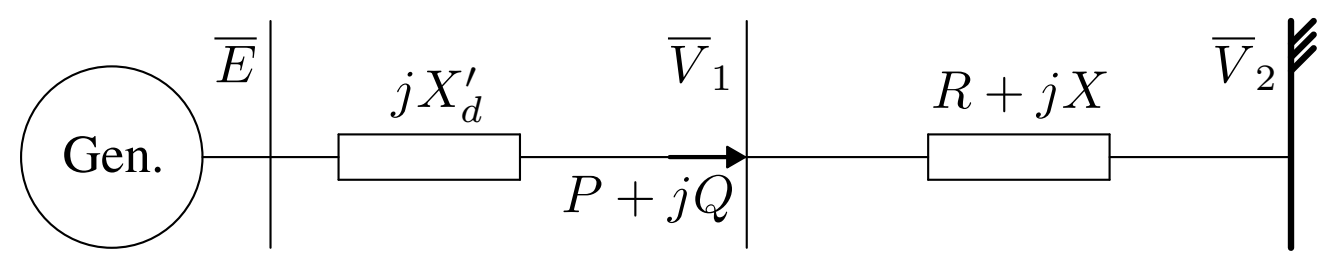
\begin{figure}[!t]
\centering
\begin{tikzpicture}[line cap=round,line join=round,>=triangle 45,x=1cm,y=1cm,scale=1.5]
\clip(8.3,3.55) rectangle (14.2,4.7);
\draw [line width=0.4pt] (8.8,4) circle (0.4cm);
\draw [line width=0.4pt] (9.5,4.6)-- (9.5,3.6);
\draw [line width=0.4pt] (11.6,4.6)-- (11.6,3.6);
\draw [line width=1.2pt] (14,4.6)-- (14,3.6);
\draw [line width=0.4pt] (13.2,4)-- (14,4);
\draw [line width=0.4pt] (12.4,4.1)-- (12.4,3.9);
\draw [line width=0.4pt] (12.4,3.9)-- (13.2,3.9);
\draw [line width=0.4pt] (13.2,3.9)-- (13.2,4.1);
\draw [line width=0.4pt] (13.2,4.1)-- (12.4,4.1);
\draw (9.17,4.62) node[anchor=north west] {$\overline{E}$};
\draw (11.17,4.62) node[anchor=north west] {$\overline{V}_{1}$};
\draw (13.57,4.62) node[anchor=north west] {$\overline{V}_{2}$};
\draw (9.95,4.48) node[anchor=north west] {$jX^{\prime}_{d}$};
\draw (12.33,4.46) node[anchor=north west] {$R+jX$};
\draw (10.7,4.00) node[anchor=north west] {$P+jQ$};
\draw (8.5,4.176831819712612) node[anchor=north west] {Gen.};
\draw [line width=1.2pt] (14.1,4.48)-- (14,4.38);
\draw [line width=1.2pt] (14.1,4.6)-- (14,4.5);
\draw [line width=1.2pt] (14.1,4.54)-- (14,4.44);
\draw [line width=0.4pt] (9.8,4.1)-- (9.8,3.9);
\draw [line width=0.4pt] (9.8,3.9)-- (10.6,3.9);
\draw [line width=0.4pt] (10.6,3.9)-- (10.6,4.1);
\draw [line width=0.4pt] (10.6,4.1)-- (9.8,4.1);
\draw [line width=0.4pt] (9.2,4)-- (9.8,4);
\draw [line width=0.4pt] (10.6,4)-- (12.4,4);
\draw [line width=0.8pt] (11.26,4)-- (11.54,4);
\begin{scriptsize}
\draw [fill=black,shift={(11.54,4)},rotate=270] (0,0) ++(0 pt,1.5pt) -- ++(1.299038105676658pt,-2.25pt)--++(-2.598076211353316pt,0 pt) -- ++(1.299038105676658pt,2.25pt);
\end{scriptsize}
\end{tikzpicture}
\caption{One-line diagram of the single-machine infinite-bus system from \cite{Netto2021}. $\overline{V}_{1}$, $\overline{V}_{2}$ are complex-valued voltage phasors in nodes $1$ and $2$, respectively. $P$ ($Q$) is the real (reactive) power injected into node $1$. $R$, $X$, and $X_{d}^{\prime}$ are parameters. We choose $R=0.05$, $X=0.30$, $V_{1}=1.05$, $V_{2}=1.00$, $P=0.80$, and $X^{\prime}_{d}=0.20$, all values in pu.}
\label{fig.smib}
\end{figure}

Our test system (Fig.~\ref{fig.smib}) is a classical generator model, parameterized with $H\!=\!5$~s, $D\!=\!10$~pu, connected to an infinite bus through a series impedance. We do not neglect the resistance $R$, which causes classical energy functions\footnote{The energy function is a specific Lyapunov-like function with physical meaning; that is, the total transient energy of the post-fault power system.} to break down. With $R>0$, path-dependent terms have no closed form \cite{Llamas1992}, and Chiang \cite{Chiang1989b} showed that no general energy function exists for lossy multimachine systems. Conversely, $R$ does not affect our proposed method. With the state $\bx=[\delta\;\omega]^{\mathsf{T}}$ shifted so that the stable equilibrium is at the origin (that is, $\bx^\star \!= \mathbf{0}$),
\begin{equation}\arraycolsep=1.0pt
\left[\begin{array}{c}
\dot\delta \\ \dot\omega
\end{array}\right] \!=\!
\left[\begin{array}{c}
\omega \\ c_{00}+c_0\omega+c_1\cos\delta+c_2\sin\delta
\end{array}\right],
\label{eq.smib}
\end{equation}
$c_{00}\!=\!21.3649$, $c_0\!=\!-1$, $c_1\!=\!-21.3649$, and $c_2\!=\!-78.5772$. See \cite{Netto2021} for details on \eqref{eq.smib}. The Jacobian at the origin, $\boldsymbol{J}(\bx^\star) =\left[\begin{smallmatrix}
0 & 1 \\ c_2 & c_0 \end{smallmatrix}\right]$, has $\lambda_{1,2}\!=\!-0.5000\pm j8.8503$. The accelerating torque of \eqref{eq.smib} at $\omega=0$,
\begin{equation}
P(\delta)=c_{00}+c_1\cos\delta+c_2\sin\delta=c_{00} + \mathrm{m}\cos(\delta-\mathrm{a}), \label{eq.power-angle}
\end{equation}
$\mathrm{m}=\sqrt{c_1^2+c_2^2}=81.4299$ and $\mathrm{a}=\atantwo(c_2,c_1)=-1.8363$. Recall that \eqref{eq.power-angle} is the familiar power-angle equation \cite[\S16.4]{Grainger1994}, but with a lossy line ($R> 0$) offset. $P(\delta)=0$ if and only if 
\begin{equation}
\cos(\delta-\mathrm{a})=-c_{00}/\mathrm{m}. \label{eq.8}
\end{equation}
Since $|-c_{00}/\mathrm{m}|=0.2624<1$, the cosine satisfies \eqref{eq.8} at two points of each period, symmetric about the maximum angle $\delta_{\max}=\mathrm{a}$. From \eqref{eq.8},
$\delta=\mathrm{a}\pm\cos^{-1}\!\left(-c_{00}/\mathrm{m}\right)=\mathrm{a}\pm\mathrm{b}$, where $\mathrm{b}\!=\!\cos^{-1}\!\left(-c_{00}/\mathrm{m}\right)\in(0,\pi)$. The equilibrium at $\delta^{\star}\!=\!\mathrm{a}+\mathrm{b}$ has $P^{\prime}(\delta^{\star})\!=\!-\mathrm{m}\sin(\delta^{\star}-\mathrm{a})\!=\!-\mathrm{m}\sin(\mathrm{b})<0$, a restoring torque; that is the stable fixed point. At $\delta_{u}\!=\!\mathrm{a}-\mathrm{b}\!=\!-3.6725$, $P^{\prime}(\delta_{u})>0$, and the fixed point is a saddle; $\delta_u+2\pi=2.6107$ is also a saddle. The saddles at $(\delta,\omega)\!=\!(-3.6725,0)$ and $(2.6107,0)$ have unstable eigenvalue $\lambda_u\!=\!8.3785$, and their stable manifolds form the boundary of the basin of the origin \cite{Chiang1987}, \cite[\S4.2]{Llamas1992}. %We use this fact to compute the exact basin boundary in Figs. \ref{fig.efun}--\ref{fig.ext}.

\section{Eigenfunctions as functions of the state}\label{sec.efun}
% \subsection{Centering the lifted variables}\label{sec.centering}
A fixed point satisfies $S_t \bx^{\star}=\bx^{\star}$ for all $t$. If $\phi$ is an eigenfunction, $\phi(S_t \bx^{\star})=e^{\lambda t}\phi(\bx^{\star})$, so $(e^{\lambda t}-1)\phi(\bx^{\star})=0$ for all $t$, and $\phi(\bx^{\star})=0$ whenever $\lambda\ne 0$. Hence, a Koopman eigenfunction with $\lambda\neq 0$ vanishes at the equilibrium.\footnote{Eigenfunctions with $\lambda = 0$ (constants, for instance) are the only ones that may be nonzero at the equilibrium. Note that a Lyapunov candidate built from an eigenfunction needs $V(\bx^{\star})=0$, and the sublevel sets must shrink to $\bx^{\star}$ as the level goes to zero.} The approximation \eqref{eq.efun_edmd} inherits this property only if the lifted fixed point is invariant under $\boldsymbol{K}$. That is, if
\begin{equation}
\boldsymbol{K}\bg(\bx^{\star})=\bg(\bx^{\star}). \label{eq.edmd.invar} 
\end{equation}
Multiplying \eqref{eq.edmd.invar} by $\bxi_k^{\mathsf{T}}$ yields $\bxi_k^{\mathsf{T}}\boldsymbol{K}\bg(\bx^{\star}) \!=\! \bxi_k^{\mathsf{T}}\bg(\bx^{\star})$. From the EDMD matrix eigendecomposition, $\bxi_k^{\mathsf{T}}\boldsymbol{K}\bg(\bx^{\star}) \!=\! \mu_k\bxi_k^{\mathsf{T}}\bg(\bx^{\star})$. From the last two expressions, we have $(\mu_k-1)\,\bxi_k^{\mathsf{T}}\bg(\bx^{\star})=0$, and from \eqref{eq.efun_edmd}, $(\mu_k-1)\,\hat\phi_k(\bx^{\star})=0$, with $\hat\phi_k(\bx^{\star})=0$ for every $\mu_k\neq1$. The least-squares solution \eqref{eq.edmd} satisfies \eqref{eq.edmd.invar} trivially when $\bg(\bx^{\star})=\boldsymbol{0}$, which holds for monomials in $(\delta,\omega)$ and for \eqref{eq.lie} below, but only approximately otherwise. For example, if $\bg_{\text{ex}}\!=\![\,\delta\ \omega\ \sin\delta\ \cos\delta\ \omega\sin\delta\ \omega\cos\delta\,]^{\mathsf{T}}$ \cite[Eq. (13)]{Netto2021}, then $\bg_{\text{ex}}(\mathbf{0})=[\,0\ 0\ 0\ 1\ 0\ 0\,]^{\mathsf{T}}$, and the spectrum contains a spurious eigenvalue $0.0529$. Centering resolves the issue at no cost, and
\begin{equation}
\bg_{\rm Lie}=[\,\delta\;\ \omega\;\ \sin\delta\;\ (\cos\delta-1)\;\ \omega\sin\delta\;\ \omega(\cos\delta-1)\,]^{\mathsf{T}}
\label{eq.lie}
\end{equation}
spans, up to the constant function, the same space as $\bg_{\text{ex}}$, and the spurious eigenvalue disappears.
%The principal pair moves from $-0.5134\pm j8.7623$ to $-0.5182\pm j8.7347$.

\begin{table}[!t]
\caption{Dictionaries considered in this letter}
\label{tab.dic}
\centering
\footnotesize
\setlength{\tabcolsep}{3pt}
\begin{tabular}{ll}
\toprule
Dictionary & Description \\
\midrule
EDMD-p$\chi$ & All monomials in $(\delta,\omega)$ of degree 1 to $\chi\in\{2,4,6\}$ \\
EDMD-Lie & The dictionary in \eqref{eq.lie}, which is adapted from \cite{Netto2021} \\
EDMD-Lie-p2 & All monomials of degree one and two in $(z_1,...,z_4)$ \\
\bottomrule
\end{tabular}
\end{table}

\subsection{Dictionaries of observable functions}
We consider five different dictionaries (Table \ref{tab.dic}). All of them are sets of monomials in the lifted variables $z_1=\delta$, $z_2=\omega$, $z_3=\sin\delta$ and $z_4=\cos\delta-1$, that is,
\begin{equation}
g_i(\bx)=\delta^{\alpha_1}\,\omega^{\alpha_2}\,(\sin\delta)^{\alpha_3}\,(\cos\delta-1)^{\alpha_4},
\label{eq.monomial}
\end{equation}
%Note that $(\sin\delta)^2=-z_4^2-2z_4$ removes one dependent monomial in EDMD-Lie-p2.
$\{\alpha_1,...,\alpha_4\}\in\mathbb{Z}_{\ge 0}$, so that a dictionary is a $q\times 4$ integer matrix of exponents. The gradient of \eqref{eq.monomial} is explicit, so the $q\times 2$ Jacobian $\boldsymbol{J}_{\bg}(\bx)$ of every dictionary is available in closed form. Using \eqref{eq.efun_edmd},
\begin{equation}
\nabla\hat\phi_k(\bx) = \sum_{i}\xi_{k,i}\,\nabla g_{i}(\bx) = \bxi_k^{\mathsf{T}}\boldsymbol{J}_{\bg}(\bx).\label{eq.efun_grad}
\end{equation}
%The row vector $\bxi_k^{\mathsf{T}}\boldsymbol{J}_{\bg}(\bx)$ is the $1\times 2$ gradient of $\hat\phi_k$. 
By \eqref{eq.efun2}, inserting the gradient \eqref{eq.efun_grad}, the Lie derivative of an EDMD eigenfunction along the model is
\begin{equation}
\Lie\hat\phi_k(\bx)=\bxi_k^{\mathsf{T}}\boldsymbol{J}_{\bg}(\bx)\,\bff(\bx),
\label{eq.lie_efun}
\end{equation}
with $\bff$ the right-hand side of \eqref{eq.smib}. Finally, define the residual
% The same expression can be obtained from the trajectory viewpoint, that is, along $\bx(t)$,
% \begin{equation*}
% \frac{d}{dt}\hat\phi_k(\bx(t)) = \bxi_k^{\mathsf{T}}\frac{d}{dt}\bg(\bx(t)) = \bxi_k^{\mathsf{T}}\boldsymbol{J}_{\bg}(\bx)\dot{\bx} = \bxi_k^{\mathsf{T}}\boldsymbol{J}_{\bg}(\bx)\bff(\bx).
% \end{equation*}
\begin{equation}
r_k \coloneqq \Lie\hat\phi_k - \hat\lambda_k\hat\phi_k \label{eq.13}
\end{equation}
of the eigenfunction equation, so that the relative residual,
\begin{equation}
\varepsilon_{k}(\bx) = \big|r_k (\bx)\big|\big/\big|\hat\lambda_k\hat\phi_k(\bx)\big|, \qquad \hat\lambda_k\ne0,\,\hat\phi_k(\bx)\ne0,
\label{eq.residual}
\end{equation}
measures, at every point, how far $\hat\phi_k$ is from satisfying \eqref{eq.efun2}.

% We are now in the position to present the main contribution of this letter.

\section{Basin of attraction from the eigenfunctions}
\subsection{Lyapunov candidates from the EDMD spectrum}\label{sec.certified}
The first candidate is $V_1=|\hat\phi_1|$, motivated by \eqref{eq.decay}. Define $\beta(t)\coloneqq|\hat\phi_1(\bx(t))|^2$ such that $\sqrt{\beta} = |\hat\phi_1|$. Then, for $\hat\phi_1\ne0$,
\begin{equation}
\dot V_1 \!=\! \frac{d}{dt}|\hat\phi_1| \!=\! \frac{d\sqrt{(\beta)}}{d\beta}\frac{d\beta}{dt} \!=\! \frac{1}{2\sqrt{\beta}}\frac{d\beta}{dt} \!=\! \frac{1}{2|\hat\phi_1|}\frac{d}{dt}|\hat\phi_1|^2. \label{eq.18}
\end{equation}
The derivative along the true vector field $\bff$ of \eqref{eq.smib} yields
\begin{align}
\frac{d}{dt}\hat\phi_1 &= \nabla\hat\phi_1\cdot\bff = \Lie\hat\phi_1, \label{eq.19} \\
\frac{d}{dt}|\hat\phi_1|^2 &= (\Lie\hat\phi_1)\hat\phi_1^* + \hat\phi_1(\Lie\hat\phi_1^*) = 2\Real(\hat\phi_1^*\Lie\hat\phi_1). \label{eq.20}
\end{align}
Substituting \eqref{eq.20} into \eqref{eq.18}, $\dot V_1 \!=\! {\Real(\hat\phi_1^*\Lie\hat\phi_1)}/{|\hat\phi_1|}$. From \eqref{eq.13},
\begin{equation}
\dot V_1=\frac{\Real\big({\hat\phi}_1^{*}\,(\hat\lambda_1\hat\phi_1+r_1)\big)}{|\hat\phi_1|}
=\Real(\hat\lambda_1)\,|\hat\phi_1|+\frac{\Real({\hat\phi}_1^{*}\, r_1)}{|\hat\phi_1|}.
\label{eq.vdot}
\end{equation}
Write $\hat\phi_1\!=\!|\hat\phi_1|e^{j\theta}$, with $e^{j\theta}$ the unit complex number pointing in the direction of $\hat\phi_1$. Then, $\hat\phi_1^*/|\hat\phi_1|=e^{-j\theta}$, the residual term in \eqref{eq.vdot} becomes $\Real({\hat\phi}_1^{*}\, r_1)/|\hat\phi_1| = \Real(e^{-j\theta}\,r_1) \eqqcolon r_{1\parallel}$, and
\begin{equation}
\dot V_1= \Real(\hat\lambda_1)\,|\hat\phi_1| + r_{1\parallel}, \label{eq.vdot2}
\end{equation}
where $r_{1\parallel}$ is the component of the residual along $\hat\phi_1$.

\noindent
The second candidate comes from a different eigenpair of $\boldsymbol{K}$. Let $\mu_E$ be the real eigenvalue of $\boldsymbol{K}$ whose continuous-time counterpart $\hat\lambda_E=\ln(\mu_E)/\Delta t$ is closest to $2\Real(\lambda_1)\!=\!-1$, with left eigenvector $\boldsymbol{\xi}_E^{\mathsf{T}}$, and let $\hat\phi_E=\boldsymbol{\xi}_E^{\mathsf{T}}\,\bg(\bx)$, the EDMD counterpart of $|\phi_1|^2$. Because $\boldsymbol{\xi}_E^{\mathsf{T}}$ is defined only up to sign, we set $V_E\!=\!s\,\hat\phi_E$, with $s=\sgn\hat\phi_E(\bx_0)$ for a fixed reference point $\bx_0$ of the training lattice; $s$ is a single constant that makes $V_E(\bx_0)>0$. From \eqref{eq.13} and \eqref{eq.19}, $\dot{V}_E \!=\! s\frac{d}{dt}\hat\phi_E \!=\!s\,\Lie\hat\phi_E \!=\! s(\hat\lambda_E\hat\phi_E+r_E) \!=\! \hat\lambda_E(s\,\hat\phi_E)+s\,r_E \!=\!\hat\lambda_EV_E+s\,r_E$.

% The second candidate is the real eigenfunction $\hat\phi_E=\hat\phi_1\hat\phi_1^*$ whose eigenvalue $\hat\lambda_E$ is closest to $2\Real(\lambda_1)\!=\!-1$, the EDMD counterpart of $|\phi_1|^2$, with $V_E\!=\!s\,\hat\phi_E$, $s=\{-1,1\}$ fixed so that $V_E>0$ near the origin. This candidate exists only if the dictionary contains that mode, which requires $\omega^2$ in its span, and according to \eqref{eq.13} and \eqref{eq.19} it satisfies $\dot{V}_E = \hat\lambda_EV_E+s\,r_E$.
\begin{lemma}
Let $\Real(\hat\lambda_1)<0$ and $\hat\lambda_E<0$. At every $\bx$ with $\hat\phi_1(\bx)\ne0$, $\dot{V}_1(\bx)<0$ if $\varepsilon_{1}(\bx)<-\Real(\hat\lambda_1)/|\hat\lambda_1|$. At every $\bx$ with $V_E(\bx)>0$, $\dot{V}_E(\bx)<0$ if $\varepsilon_{E}(\bx)<1$.
\end{lemma}
\begin{proof}
$|r_{1\parallel}|\!=\!|\Real(e^{-j\theta}\,r_1)| \!\le\! |e^{-j\theta}\,r_1|\!=\!|e^{-j\theta}|\,|r_1|\!=\!|r_1|$. Thus, $\dot{V}_1\!=\!\Real(\hat\lambda_1)\,|\hat\phi_1| \!+\! r_{1\parallel}\!\le\!-|\Real(\hat\lambda_1)|\,|\hat\phi_1|\!+\!|r_1|\!=\!-|\Real(\hat\lambda_1)|\,V_1\,(1\!-\!\varepsilon_{1}|\hat\lambda_1|/|\Real(\hat\lambda_1)|)$. The prefactor $-|\Real(\hat\lambda_1)|\,V_1$ is strictly negative, so $\dot V_1\!<\!0$ whenever $(1\!-\!\varepsilon_{1}|\hat\lambda_1|/|\Real(\hat\lambda_1)|)\!>\!0$, that is, whenever $\varepsilon_{1}\!<\!-\Real(\hat\lambda_1)/|\hat\lambda_1|$. For the second candidate, $\dot{V}_{E}\!\le\!-|\hat\lambda_E|\,V_E\,(1\!-\!\varepsilon_{E})$.
\end{proof}
For our system \eqref{eq.smib}, $-\Real(\hat\lambda_1)/|\hat\lambda_1|\!\approx\!0.056$ (e.g., $0.059$ for EDMD-Lie), and the modulus of the principal eigenfunction decreases wherever the relative residual is below $6\%$. For the second candidate, the real eigenfunction decreases wherever the residual is below $100\%$. The difference is the eigenvalue. For a complex pair, decay is a small fraction of the rotation rate, and only the decay competes with the residual. For a real negative eigenvalue, the whole of $|\hat\lambda_E|$ does. Thus, $V_E$ certifies larger sets than $V_1$ in Section \ref{sec.numres}, and it suggests a design rule: real eigenfunctions with eigenvalues $2p\Real(\lambda_1)$, $p\ge1$, are the natural Lyapunov candidates in the EDMD spectrum.
\begin{remark}\label{rem.V1V}
Let $V=|\hat\phi_1|^2$. For any $\bx$ and any $c\ge0$, $|\hat\phi_1|\le c$ if and only if $|\hat\phi_1|^2\le c^2$, since squaring is strictly increasing on $[0,\infty)$, and $|\hat\phi_1|$ takes values there. Thus, the set $\{\bx:V_1(\bx)\le c\}$ is the same set of points as $\{\bx:V(\bx)\le c^2\}$. $V$ and $V_1$ yield the same family of sublevel sets; only the number attached to each set changes, from $c$ to $c^2$. Both sets have the same boundaries. Since squaring does not change the sublevel sets, $|\hat\phi_1|^2$ certifies exactly the sets of $V_1$ in Table \ref{tab.res}. Therefore, the gap between the $V_1$ and $V_E$ columns measures the EDMD error in the complex mode, not the candidate shape.
\end{remark}

\subsection{Certified sublevel sets}
\begin{proposition}[Sublevel-set certificate {\cite[\S8.2]{Khalil2002}}]\label{prop.sublevel}
Let $V$ $\in C^1(W)$ on an open set $W\ni\mathbf{0}$ with $V(\mathbf{0})=0$, and let $\Omega_c$ be the connected component of $\{\bx\in W:V(\bx)\le c\}$ that contains the origin. If $\Omega_c$ is compact, $V>0$ and $\dot V<0$ on $\Omega_c\setminus\{\mathbf{0}\}$, then $\Omega_c$ is positively invariant and $\Omega_c\subset\mathcal{B}(\bx^\star\!=\,\mathbf{0})$.
\end{proposition}

\begin{figure}[t!]
\centering
\includegraphics[width=\columnwidth]{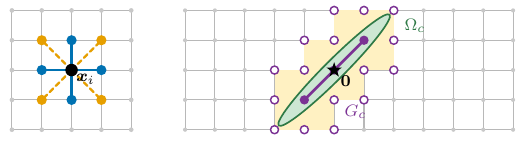}
\caption{\emph{Left}. A grid point $\bx_i$ and its eight neighbors, the four that share an edge of a cell with it (solid) and the four that share only a corner (dashed). \emph{Right}. A connected component $\Omega_c$ of a sublevel set (green) running along the diagonal of the lattice. Filled markers are grid points inside $\Omega_c$. Shaded cells meet $\Omega_c$; consecutive ones share a corner, and every corner of every shaded cell (filled and hollow markers) satisfies $V\!\le\! c+2L_V\rho$ and belongs to the collar component $G_c$ of Lemma \ref{lem.grid}.}
\label{fig.adj}
\end{figure}

$V_1$ is not differentiable at the origin, so we apply Proposition \ref{prop.sublevel} to $V$, whose sublevel set at level $c^2$ is $\Omega_c$ (cf. Remark \ref{rem.V1V}). Both conditions of Proposition \ref{prop.sublevel} are monotone in $c$ because sublevel sets nest, so the largest certified level $c^\star$ is found by bisection. We evaluate $V$ and $\dot V$ from \eqref{eq.efun_edmd} and \eqref{eq.lie_efun} on the lattice $\delta\in[-4,4]$, $\omega\in[-20,18]$ with steps $0.025$~rad and $0.1$~rad/s. We compute $\Omega_c$ from a flood fill\footnote{Starting from a seed cell (the origin, in our case) of a grid, the algorithm marks the seed, then marks every 8-neighbor of a marked cell that satisfies a membership test, and repeats until no new cell can be added \cite[\S9.5.3]{Gonzalez2008}.} from the origin, and reject a level if the component touches the border of the window or if any of its grid points other than the origin violates a condition. Let $\rho$ be the half-diagonal of a lattice cell, so that every point of $W$ lies within $\rho$ of a grid point (Fig. \ref{fig.adj}). For a level $c$, let $G_c$ be the set of grid points reachable from the origin through 8-adjacent grid points $\bx_i$ with $V(\bx_i)\le c+2L_V\rho$, $L_V\in\mathbb{R}$.
% \begin{remark}[Rejecting components that touch the border] A component touches the border if it contains a grid point in the first or last row or column of the lattice. Proposition \ref{prop.sublevel} requires $\Omega_c$ to be compact and contained in $W$, but a component that reaches the window's edge may extend beyond the window. Therefore, the lattice truncates the true set rather than representing it.
% \end{remark}
\begin{lemma}[From a discrete lattice to the continuum \cite{Cavraro2026}]\label{lem.grid}
Let $V \in C^1(W)$ with $V(\mathbf{0})=0$, and let $V$ and $\dot V$ be Lipschitz on $W$ with constants $L_V$ and $L_{\dot V}$. Suppose that
\begin{enumerate}
\item[(a)] $V>0$ and $\dot V<0$ on $B_{r_0}(\mathbf{0})\setminus \{\mathbf{0}\}$, an open disk of radius $r_0>\rho$;
\item[(b)] every $\bx_i\in G_c$ with $|\bx_i|\ge r_0-\rho$ satisfies $V(\bx_i)> L_V\rho$ and $\dot V(\bx_i)\le -\eta$, with $\eta> L_{\dot V}\rho$;
\item[(c)] no point of $G_c$ lies on the border of the lattice.
\end{enumerate}
Then, $\Omega_c$ satisfies Proposition \ref{prop.sublevel}, and $\Omega_c\subset\mathcal{B}(\mathbf{0})$.
\end{lemma}
\begin{proof}
Let $\mathcal{Q}$ be the set of closed lattice cells that meet $\Omega_c$. If $\mathcal{Q}$ could be split into two families such that no cell of one touches a cell of the other, the unions of the two families would be disjoint closed sets covering $\Omega_c$, contradicting its connectedness; hence any two cells of $\mathcal{Q}$ are linked by a chain of cells in $\mathcal{Q}$ in which consecutive cells share a corner. Every cell of $\mathcal{Q}$ contains a point with $V\le c$, and each of its corners lies within $2\rho$ of that point, so every corner of every cell of $\mathcal{Q}$ satisfies $V\le c+2L_V\rho$. The origin is such a corner, so all these corners belong to $G_c$. By (c), no cell of $\mathcal{Q}$ is an outermost cell. Thus, $\Omega_c$ lies in the closed inner rectangle and, being closed in $W$, it is compact. Now, let $\bx\in\Omega_c\setminus\{\mathbf{0}\}$. If $|\bx|<r_0$, (a) applies. Otherwise, let $\bx_i$ be the corner nearest to $\bx$ of a cell of $\mathcal{Q}$ containing $\bx$. Then, $|\bx-\bx_i|\le\rho$, $\bx_i\in G_c$, and $|\bx_i|\ge|\bx|-\rho\ge r_0-\rho$, so (b) and Lipschitz continuity yield $V(\bx) \!\ge\! V(\bx_i)-L_V\rho>0$ and $\dot V(\bx)\le\dot V(\bx_i)+L_{\dot V}\rho<0$.
\end{proof}
\begin{remark}[Families]
Positively invariant subsets of the basin are closed under union, so the union of certified sets over both candidates and all dictionaries is itself certified, as in the Lyapunov-function-family approach of \cite{Vu2016}.
\end{remark}
\begin{remark}
Lemma \ref{lem.grid} gives the conditions under which the grid test of this subsection certifies the continuum. We evaluate the extension along the flow in \S\ref{sec.extension} at the grid points by numerical integration, and the basin estimate it produces holds at the grid resolution, up to integration error.
\end{remark}

\subsection{Extension along the flow}\label{sec.extension}
The certified set is small when the training data are confined to the linear region because $\hat\phi_1$ is accurate only near the origin. The semigroup property repairs this \emph{without new data}, at the cost of numerical integration. This matters because power systems have multiple attractors, operate near a stable fixed point most of the time, and one seeks to estimate the basin from measured ring-downs after small disturbances. Let $\Omega_0\subseteq\Omega_{c^\star}$ be the connected component of $\{\bx:|\hat\phi_1(\bx)|\le c_0\}$ that contains the origin, $0<c_0\le c^\star$, chosen so that $\varepsilon_1\le\varepsilon_{\rm tol}$ on $\Omega_0$. Define 
\begin{equation}
\tau(\bx)\coloneqq \inf\{t\ge 0: S_t\bx\in\Omega_0\}\in[0,\infty], \label{eq.tau}
\end{equation}
the hitting time of the seed. Since $S_t\bx$ is the state at time $t$ of the trajectory starting at $\bx$, the set $\{t\!\ge\! 0: S_t\bx\in\Omega_0\}$ collects all times when that trajectory is in the seed set $\Omega_0$. The infimum of the set in \eqref{eq.tau} therefore defines, for each state $\bx$, the first time the trajectory starting at $\bx$ enters $\Omega_0$. Define
\begin{equation}
\tilde\phi_1(\bx)\coloneqq e^{-\hat\lambda_1\tau(\bx)}\,\hat\phi_1\left(S_{\tau(\bx)}\bx\right) \qquad \text{on}\, \{\tau<\infty\}.
\label{eq.ext}
\end{equation}
\begin{proposition}[Properties of the extension]
Let $\Omega_0$ and $\tilde\phi_1$ be as before. Then:
\begin{enumerate}
\item[(i)] $\Omega_0$ is compact and positively invariant, contains a neighborhood of the origin, and $\{\tau<\infty\}=\mathcal{B}(\mathbf{0})$.
\item[(ii)] For every $T\ge0$, $\{\tau\le T\}=S_{-T}(\Omega_0)$. These sets are compact, positively invariant, nested in $T$, and their union is $\mathcal{B}(\mathbf{0})$. The hitting time is continuous on $\mathcal{B}(\mathbf{0})$, continuously differentiable on $\mathcal{B}(\mathbf{0})\setminus\Omega_0$ with $\dot\tau=-1$ there, and $\tau(\bx)\to\infty$ as $\bx\to\partial\mathcal{B}(\mathbf{0})$.
% \item[(iii)]
\end{enumerate}
\end{proposition}
\begin{proof}
(i) Proposition \ref{prop.sublevel} applied to $V=V_1^2$ at level $c_0^2$ yields compactness and positive invariance, since $V_1>0$ and $\dot V_1<0$ on $\Omega_{c^\star}\setminus\{\mathbf{0}\}\supseteq\Omega_0\setminus\{\mathbf{0}\}$. The open set $\{V_1<c_0\}$ contains the origin, and its component through the origin lies in $\Omega_0$. If $\tau(\bx)<\infty$, the infimum in \eqref{eq.tau} is attained because $\Omega_0$ is closed, so $S_{\tau(\bx)}\bx\in\Omega_0\subseteq\mathcal{B}(\mathbf{0})$ and $\bx\in\mathcal{B}(\mathbf{0})$; conversely, $S_t\bx\to\mathbf{0}$ for $\bx\in\mathcal{B}(\mathbf{0})$, so $S_t\bx\in\Omega_0$ for large $t$. 
(ii) $\tau(\bx)\le T$ if and only if $S_T\bx\in\Omega_0$: the ``if'' is the definition of $\tau$, the ``only if'' is positive invariance of $\Omega_0$. The vector field of \eqref{eq.smib} is globally Lipschitz (its Jacobian is bounded), so its flow is complete \cite[Thm.~3.2]{Khalil2002}, $S_{-T}$ is a homeomorphism of $\mathbb{R}^2$, and $\{\tau\le T\}=S_{-T}(\Omega_0)$ is compact. Positive invariance of $\Omega_0$ means $\Omega_0\subseteq S_{-s}(\Omega_0)$ for $s\ge 0$; applying $S_{-T}$ yields $S_{-T}(\Omega_0)\subseteq S_{-(T+s)}(\Omega_0)$, so the sets are nested in $T$. For $0\le t\le T$, $S_t\big(S_{-T}(\Omega_0)\big)=S_{-(T-t)}(\Omega_0)$, and for $t>T$ it equals $S_{t-T}(\Omega_0)\subseteq\Omega_0$; both lie in $S_{-T}(\Omega_0)$ by nestedness, so each set is positively invariant. Their union is $\{\tau<\infty\}=\mathcal{B}(\mathbf{0})$ by (i). Since $\dot V_{1}<0$ on $\partial\Omega_0$, every trajectory from $\mathcal{B}(\mathbf{0})\setminus\Omega_0$ enters $\Omega_0$ transversally through $\partial\Omega_0$, and by continuous dependence on initial conditions nearby trajectories enter at nearby times; the implicit function theorem applied to $V_1(S_t\bx)=c_0$ at the entry time then gives that $\tau$ is $C^1$ on $\mathcal{B}(\mathbf{0})\setminus\Omega_0$, and the same argument provides continuity on $\mathcal{B}(\mathbf{0})$. Differentiating $\tau(S_t\bx)=\tau(\bx)-t$ at $t=0$ yields $\dot\tau=-1$. Finally, for every $T$, the set $\{\tau\le T\}$ is a compact subset of the open set $\mathcal{B}(\mathbf{0})$, so a sequence approaching $\partial\mathcal{B}(\mathbf{0})$ eventually leaves each of them, and $\tau\to\infty$.
\end{proof}

Since $\tau(S_t\bx)=\tau(\bx)-t$ for $t\le\tau(\bx)$, \eqref{eq.ext} satisfies $\tilde\phi_1(S_t\bx)=e^{\hat\lambda_1 t}\tilde\phi_1(\bx)$ outside $\Omega_0$. If $\phi_1$ is the exact eigenfunction, the semigroup property yields $\phi_1(\bx)/\tilde\phi_1(\bx) = \phi_1(\boldsymbol{y})/\tilde\phi_1(\boldsymbol{y})$ at the entry point $\boldsymbol{y}=S_{\tau(\bx)}\bx$, with the seed error carried along the flow unchanged, while the error of the eigenvalue enters through $e^{(\hat\lambda_1-\lambda_1)\tau(\bx)}$. This is the mechanism behind Laplace averages \cite{Mauroy2016} and the construction of \cite{Korda2020}, with a finite integration time because EDMD supplies an accurate seed. The identity $\{\tau\le T\}=S_{-T}(\Omega_0)$ is the trajectory-reversing construction of \cite{Genesio1985}. EDMD yields a larger seed, hence a shorter horizon, and isostables and isochrons on $\Omega_0$ without integration. A trajectory that starts at a small distance $d>0$ from the basin boundary, that is, from the stable manifold of a saddle, is first drawn toward the saddle and only then expelled along its unstable direction, so it lingers near the saddle for a time that grows like $\lambda_u^{-1}\ln(1/d)$ before heading to the seed. The hitting time is therefore $\tau=\lambda_u^{-1}\ln(1/d)+O(1)$ as $d\to0$. Since the trajectory enters the seed through $\partial\Omega_0$, where $|\hat\phi_1|=c_0$, \eqref{eq.ext} yields $|\tilde\phi_1(\bx)|=c_0\, e^{-\Real(\hat\lambda_1)\tau(\bx)}$ outside $\Omega_0$, and substituting the hitting time, $|\tilde\phi_1(\bx)|= \kappa\,e^{-\Real(\hat\lambda_1)\lambda_u^{-1}\ln(1/d)} = \kappa\,d^{\Real(\hat\lambda_1)/\lambda_u}$, with $\kappa$ bounded away from zero and infinity. The exponent is negative because $\Real(\hat\lambda_1)<0$, so the modulus grows without bound as we approach the boundary. This is the mechanism by which the eigenfunction fails to extend beyond the basin \cite{Mauroy2016}. The level sets of $|\tilde\phi_1|$, the sets $S_{-T}(\partial\Omega_0)$, are the isostables, and those of $\arg\tilde\phi_1$ the isochrons \cite{Mauroy2013}.

\section{Numerical results}\label{sec.numres}
The training data comprise 45 trajectories with initial states on the lattice $\delta\in\{-0.50{:}0.25{:}0.50\}$, $\omega\in\{-1.00{:}0.25{:}1.00\}$, sampled every $\Delta t=5$~ms for 0.8~s (7200 snapshot pairs). Note that the training data come from a linear region of the state space near the stable equilibrium point.

\begin{table}[!t]
\caption{EDMD eigenvalues and fraction of the basin certified by the sublevel sets of $V_1$ and $V_E$}
\label{tab.res}
\centering
\begin{threeparttable}
\footnotesize
\setlength{\tabcolsep}{4pt}
\begin{tabular}{lrcrrr}
\toprule
Dictionary & $q$ & $\hat\lambda_1$ & $\hat\lambda_E$ & $V_1$ & $V_E$ \\
\midrule
linearization & 2  & $-0.5000\pm j8.8503$ & ---       &  8.3\% & ---    \\
EDMD-p2       & 5  & $-0.5011\pm j8.7504$ & $-1.0048$ & 23.9\% &  0.0\% \\
EDMD-p4       & 14 & $-0.4949\pm j8.8487$ & $-1.0001$ & 34.3\% & 60.0\% \\
EDMD-p6       & 27 & $-0.5000\pm j8.8504$ & $-1.0000$ & 55.9\% & 66.1\% \\
EDMD-Lie      & 6  & $-0.5182\pm j8.7347$ & absent    &  0.2\% & ---    \\
EDMD-Lie-p2   & 13$^\ddag$ & $-0.5153\pm j8.7655$ & $-0.9930$ &  6.9\% & 16.5\% \\
\bottomrule
\end{tabular}
\begin{tablenotes}
\item[]$^\ddag13$ after removing $z_3^2=-z_4^2-2z_4$.
\end{tablenotes}
\end{threeparttable}
\end{table}

\begin{figure}[!t]
\centering
\includegraphics[width=0.49\textwidth]{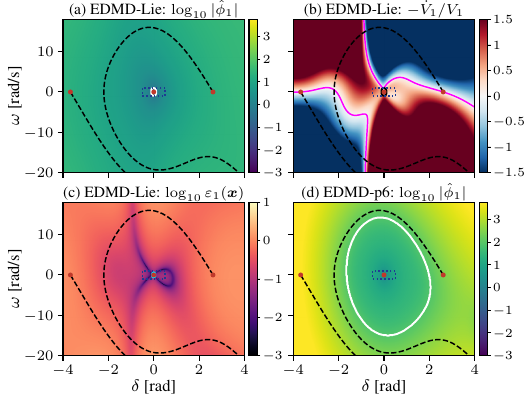}
\caption{EDMD eigenfunctions evaluated on the state space. (a) $\log_{10}|\hat\phi_1|$ for EDMD-Lie, with the certified sublevel set of $V_1$ in white. (b) Local decay rate $-\dot V_1/V_1$ of the EDMD-Lie eigenfunction along the true vector field (exact value $0.5$), with $\dot V_1=0$ in magenta and the certified set in black. (c) Relative residual \eqref{eq.residual} of EDMD-Lie, with the seed set $\Omega_0$ in cyan. (d) $\log_{10}|\hat\phi_1|$ for EDMD-p6 with its certified set. In all panels, red dots are fixed points; the dotted rectangle is the training lattice; the black dashed line is the stable manifold of the saddles (exact basin boundary), computed by integrating \eqref{eq.smib} backward in time from the saddles along their stable eigenvectors \cite{Chiang1987}.}
\label{fig.efun}
\end{figure}

% \begin{figure*}[!t]
% \centering
% \includegraphics[width=.75\textwidth]{fig_roa.pdf}
% \caption{Certified sublevel sets (inner estimates of the basin) obtained with (a) $V_1=|\hat\phi_1|$ and (b) $V_E=s\,\hat\phi_E$ for the five dictionaries; legends give the fraction of the basin covered. Gray: true basin inside the window; dashed: stable manifolds of the saddles; dotted rectangle: training lattice.}
% \label{fig.roa}
% \end{figure*}

\begin{figure}[!t]
\centering
\includegraphics[width=.97\linewidth]{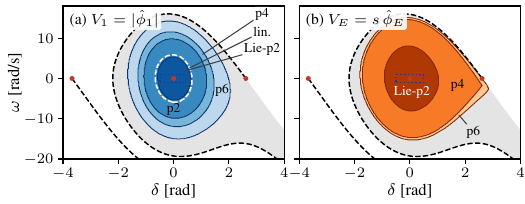}
\caption{Certified sublevel sets of (a) $V_1$ and (b) $V_E$, drawn as nested fills over the true basin (gray) and labeled by the dictionary in Table \ref{tab.res}. The dashed white line in (a) is the linearization.}
\label{fig.roa}
\end{figure}

We compare the results obtained with the five dictionaries in Table \ref{tab.dic}, along with the linearization ($\hat\phi_1=\bl_1^{\mathsf{T}}\bx_0$). Table~\ref{tab.res} lists the principal eigenvalues recovered by all dictionaries within 1.3\% in frequency and 3.6\% in decay rate, and the real eigenvalue $\hat\lambda_E$ closest to $2\Real(\lambda_1)$, which appears as soon as $\omega^2$ is in the span. Note that the six-function Lie dictionary does not contain it, its closest real eigenvalue being $-2.835$.

% \subsection{Eigenfunctions and their validity}
Fig.~\ref{fig.efun}(a) shows the modulus of the EDMD-Lie principal eigenfunction and Fig.~\ref{fig.efun}(b) shows the local decay rate of $|\hat\phi_1|$ along the true flow. Note that the exact eigenfunction would give the constant $0.5$ on the whole basin, whereas the EDMD eigenfunction gives it in a cross-shaped region around the axes and fails, with the wrong sign, in the first and third quadrants beyond $|\delta|\approx 0.3$ and $|\omega|\approx 4$. This is because the true isostables follow, to first approximation, the level sets of the transient energy $\tfrac12\omega^2+U(\delta)$, whereas \eqref{eq.lie} is linear in $\omega$. Note that $U(\delta)$ is the potential energy of \eqref{eq.smib} \cite{Llamas1992}. The Lie derivative of $\omega(\cos\delta-1)$ contains $-\omega^2\sin\delta$, which has no counterpart in the span, and the corresponding residual term grows as $\omega^2\delta$. Monomials of degree six, Fig.~\ref{fig.efun}(d), reproduce the eigenfunction over a much larger region because the lattice data identify the Taylor coefficients of $\phi_1$ to that order.

% \begin{figure*}[!t]
% \centering
% \includegraphics[width=.80\textwidth]{fig_ext.pdf}
% \caption{Extension \eqref{eq.ext} of the EDMD-Lie eigenfunction. (a) $|\tilde\phi_1|$ (isostables); white: levels reached at hitting times $\tau=0,1,\dots,8$~s. (b) $\arg\tilde\phi_1$ (isochrons). (c) Basin estimate as the domain where $\tilde\phi_1$ is finite (blue), true basin (gray) and stable manifolds (dashed); colored curves: isostables at $\tau=0,\dots,5$~s.}
% \label{fig.ext}
% \end{figure*}

\begin{figure}[!t]
\centering
\includegraphics[width=.98\linewidth]{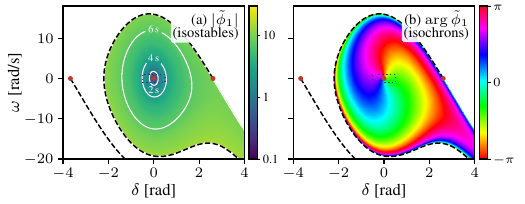}
\caption{Extension \eqref{eq.ext} of the EDMD-Lie eigenfunction. (a) $|\tilde\phi_1|$ on a logarithmic scale. The white lines are the isostables labeled by their hitting time $\tau=2,4,6$ s ($\tau=8.09$ s lies within the basin-boundary line width). (b) $\arg\tilde\phi_1$ (isochrons). In both, the colored region is the basin estimate $\{\bx:\tau(\bx)<\infty\}$, which coincides with the true basin (dashed) at the grid resolution.}
\label{fig.ext}
\end{figure}

% \subsection{Certified estimates}
Table~\ref{tab.res} and Fig.~\ref{fig.roa} report the certified sublevel sets; every one is a strict subset of the true basin. With $V_1$, the estimates grow with the polynomial degree, from 8\% for the quadratic Lyapunov function of the linearization to 56\% for EDMD-p6, and the Lie dictionary is the most conservative at 0.2\%, for the reason discussed in the previous section. With $V_E$, the real eigenfunction approximates $|\phi_1|^2$, an energy-like function quadratic in the lifted variables, and it certifies 60\% and 66\% of the basin for EDMD-p4 and EDMD-p6, with a set that reaches the saddle at $(2.61,0)$. For EDMD-p2, the mode is present, but its eigenfunction carries small spurious linear terms, so it is not positive definite and certifies nothing. EDMD-Lie-p2 improves on EDMD-Lie (17\% with $V_E$) yet remains below the monomials. Indeed, on data from the linear region, $\sin\delta$, $\cos\delta-1$, and their products are nearly collinear with the monomials they approximate, and the condition number of $\boldsymbol{G}_0$ grows from $7\times10^{4}$ (EDMD-Lie) to $7\times10^{9}$ (EDMD-Lie-p2) and $6\times10^{17}$ for the degree-three lifted set, which is why the family stops at degree two.

Sampling the training trajectories over the basin (200 trajectories from $\delta\in[-2,2]$, $\omega\in[-12,12]$, 5~s each) does not help. The least-squares fit is then dominated by large-amplitude data outside the span of every dictionary. No dictionary certifies more than 0.1\% of the basin. This dictionary non-invariance \cite{Korda2018} motivates the extension in Section~\ref{sec.extension}.

% \subsection{Extension along the flow}
The seed set $|\hat\phi_1|\le c_0=0.3139$ of the EDMD-Lie eigenfunction (certified level $c^\star=0.4981$, $\varepsilon\le 5\%$) contains 54 grid points, 0.09\% of the basin. All grid points are integrated with a vectorized fourth-order Runge--Kutta scheme (step 5~ms, 40~s), recording the first entry into the seed set. Every point of the true basin, 58,654 of the 122,301 grid points, reaches it within 8.09~s and no point outside does, so the basin estimate, the colored region of Fig.~\ref{fig.ext}(a), coincides with the basin at the grid resolution. The short hitting times reflect the strong unstable eigenvalue of the saddles: trajectories that start near the stable manifolds are expelled within a fraction of a second, and $|\tilde\phi_1|$ grows by a factor of only 66 between the seed set and the basin boundary, Fig.~\ref{fig.ext}(a), consistent with the exponent $-0.062$ of Section~\ref{sec.extension}. The isochrons of Fig.~\ref{fig.ext}(b) wind around the origin with the spiral of a lightly damped focus and accumulate along the saddle manifolds.

\section{Conclusion}
EDMD eigenfunctions can be used beyond the snapshots on which they are learned once they are written as functions of the state. For the single-machine infinite-bus system, two Lyapunov candidates built from the spectrum give certified inner estimates of the basin of attraction, the real eigenfunction with eigenvalue $2\Real(\hat\lambda_1)$ reaching two-thirds of the basin with monomials of degree six learned in the linear region. The Lie-lifted observables that excel in Koopman Kalman filtering produce the smallest certificate, because the dictionary is linear in the rotor speed, and a second lifting round is limited by conditioning. The semigroup property removes the dependence of the basin estimate on the dictionary. Indeed, extended along the flow from a small certified seed, the EDMD-Lie eigenfunction is defined exactly on the basin, and the basin of attraction is recovered at the grid resolution. We show that centering the lifted variables so that the fixed point is invariant under the EDMD matrix, and verifying the decrease of the candidate with the model rather than assuming it from the spectrum, are decisive. Future work will address multimachine power systems and sum-of-squares replacements for the grid-based check.

\bibliographystyle{IEEEtran}
\bibliography{lib}

\end{document}